\pdfoutput=1

\documentclass[12pt]{article}
\usepackage[colorlinks=true, urlcolor=blue, linkcolor=blue, citecolor=blue]{hyperref} 
\usepackage{amsmath,amsthm,amsfonts,amssymb,amscd} 
\usepackage{bm}
\usepackage{subcaption} 
\usepackage{graphicx} 
\usepackage[font={small}]{caption} 
\usepackage{booktabs} 
\usepackage[authoryear]{natbib} 
\usepackage{setspace} 
\usepackage{authblk}
\usepackage{dsfont}
\usepackage{float}
\usepackage{xcolor}
\usepackage{parskip}
\usepackage{indentfirst}
\usepackage[left= 1in,right=1in,top=1in,bottom=1in]{geometry}
\usepackage{fancyhdr}
\usepackage{soul}

\newcommand{\tp}{\mathsf{T}}

\newcommand{\alp}{\mbox{\boldmath $\alpha$}}

\newcommand{\U}{\mbox{\boldmath $U$}}
\newcommand{\D}{\mbox{\boldmath $D$}}
\newcommand{\V}{\mbox{\boldmath $V$}}
\newcommand{\Z}{\mbox{\boldmath $Z$}}
\newcommand{\R}{\mbox{\boldmath $R$}}
\newcommand{\X}{\mbox{\boldmath $X$}}
\newcommand{\E}{\mbox{\boldmath $E$}}

\newcommand{\W}{\mbox{\boldmath $W$}}

\newcommand{\M}{\mbox{\boldmath $M$}}

\newcommand{\bS}{\mbox{\boldmath $S$}}

\newcommand{\B}{\mbox{\boldmath $B$}}

\newcommand{\bv}{\mbox{\boldmath $v$}}

\newcommand{\bze}{\mbox{\boldmath $0$}}

\title{Spatial Matrix Completion for Spatially-Misaligned and High-Dimensional Air Pollution Data}

\author[1]{Phuong T. Vu}
\author[1]{Adam A. Szpiro}
\author[1]{Noah Simon}
\affil[1]{Department of Biostatistics, University of Washington}
\date{}

\footnotetext{\noindent Corresponding author: Phuong T. Vu, Current Affiliation: Seattle Children's Research Institute, 1920 Terry Ave, Seattle, WA-98101. Email: phuongvu@uw.edu.}

\begin{document}

\maketitle
\doublespacing

\begin{abstract}
In health-pollution cohort studies, accurate predictions of pollutant concentrations at new locations are needed, since the locations of fixed monitoring sites and study participants are often spatially misaligned. For multi-pollution data, principal component analysis (PCA) is often incorporated to obtain low-rank (LR) structure of the data prior to spatial prediction. Recently developed predictive PCA modifies the traditional algorithm to improve the overall predictive performance by leveraging both LR and spatial structures within the data. However, predictive PCA requires complete data or an initial imputation step. Nonparametric imputation techniques without accounting for spatial information may distort the underlying structure of the data, and thus further reduce the predictive performance. We propose a convex optimization problem inspired by the LR matrix completion framework and develop a proximal algorithm to solve it. Missing data are imputed and handled concurrently within the algorithm, which eliminates the necessity of a separate imputation step. We review the connections among those existing methods developed for spatially-misaligned multivariate data, and show that our algorithm has lower computational burden and leads to reliable predictive performance as the severity of missing data increases. 

\begin{center}{\small \textbf{Keywords:} low-rank matrix completion, principal component analysis, proximal algorithm, spatial prediction}\end{center}
\end{abstract}

\newpage
\doublespacing

\section{Introduction \label{sec-intro}}

In multi-pollutant studies, a dataset is often represented as an $(n\times p)$ matrix $\X$, in which the concentrations of $p$ pollutants are collected at $n$ monitoring locations. When evaluating the associations between health outcomes and exposures to air pollution, including some or all of these pollutants in a statistical model can be problematic due to correlations and potential interactions among these components. Hence, dimension reduction is often necessary to obtain a lower-dimensional representation of the original data. 

Principal component analysis (PCA) \citep{jolliffe1986principal} is an unsupervised technique for dimension reduction that has been used in multi-pollutant analysis \citep{dominici2003health}. PCA essentially provides a mapping of the original data $\X$ to a low-rank (LR) approximation $\U \V^\tp$, in which $\U \in \mathds{R}^{n\times q}$ and $\V \in \mathds{R}^{p\times q}$ $(q < p)$ are usually referred to as principal component (PC) scores and loadings. The product $\U \V^\tp$ can be considered to be the best rank-$q$ approximation to $\X$. One can also derive this quantity by solving $\displaystyle{\min_{\W}} \hspace{0.15cm} \left\{ \frac{1}{2} \big\Vert \X - \W \big\Vert^2_F + \lambda \big\Vert \W \big\Vert_0\right\}$, for a value of $\lambda$ dependent on $q$. Here $\big\Vert \W \big\Vert_0$ denotes the number of non-zero singular values of $\W$. Another approach is to replace this L0 penalty, $\big\Vert . \big\Vert_0$, with the nuclear norm, 
\begin{align}
\min_{\W} \hspace{0.15cm} \left\{ \frac{1}{2} \big\Vert \X - \W \big\Vert^2_F + \lambda \big\Vert \W \big\Vert_*\right\}. \label{eq:tradl1}
\end{align}
Here $\big\Vert \W \big\Vert_*$ denotes the nuclear norm of $\W$, which is equal to the sum of its singular values. When closed-form solutions exist for both problems, the L0 penalty is not convex. When some elements of $\X$ are missing, the extension to matrix completion with an L0 norm is NP-hard, and existing algorithms would require time doubly exponential in the matrix's dimensions \citep{candes2009exact}. With missing data, the convex relaxation of the nuclear norm allows optimization via semidefinite programming, and leads to efficient LR matrix completion (LRMC) algorithms \citep{cai2010singular, mazumder2010spectral}. LRMC is a powerful tool to reduce dimension while being able to utilize incomplete information. 

When multiple pollutants are involved, it is often more challenging as modeling many correlated pollutant surfaces can be intractable. The goal is then to leverage both the LR and spatial structures in the data at monitoring locations in order to recover unobserved data at new locations of interest. While PCA can produce a mapping from the original data to their LR structure, LRMC can take advantage of the LR structure even when the monitoring data have missing observations. While both methods cannot, by themselves, produce estimates of the pollutant concentrations at new locations, the resulted LR structure can be incorporated in a spatial prediction model. However, these methods do not account for external geographic information and spatial correlations across neighboring locations. The lack of utilizing spatial structure in these methods may lead to poor predictive performance at new locations.

A spatially predictive PCA algorithm, PredPCA \citep{jandarov2017novel}, was developed to produce LR structure embedded with spatial patterns that can be subsequently predicted well at new locations. While capable of leveraging both the LR and spatial structures of the monitoring data, in its current form PredPCA cannot handle input data with missing observations. This motivated a probabilistic version of PredPCA, ProPrPCA \citep{vu2019probabilistic}, that can handle incomplete monitoring data. Using external covariates, the unobserved data at new locations can be estimated directly from the hierarchical model assumptions, or by incorporating the LR structure of monitoring data in a separate spatial prediction model. However, similar to PredPCA, the estimation procedure of ProPrPCA is computationally burdensome, due to a combination of multiple rank-1 approximations, numerical optimization, and likelihood-base imputations.   

In this paper, our objective is to develop a practical and computationally feasible of version of PCA that can be applied to the setting of spatially-misaligned multivariate data. Our proposed method, Spatial Matrix Completion, aims to (i) accommodate complex spatial missing patterns and (ii) incorporate spatial information into the LR structure of the data so that estimates at new locations can be easily and accurately derived. we leverage the LRMC framework to formulate a convex optimization problem, and derive a straight-forward algorithm to solve it using proximal gradient descent. While solving similar problem to the existing method ProPrPCA, our proposed method offers a simple algorithm that is easy to implement, computationally efficient, and potentially scalable. In addition, we also discuss the connections among existing methods (LRMC, PredPCA, and ProPrPCA) in conjunction with our proposed method. 

For the rest of this paper, we first review the LRMC framework in Section 2. We then propose a spatial matrix completion problem, along with its solution and connections with existing methods in Section 3. In Section 4, we compare the performance of our proposed method to LRMC in recovering missing entries at monitoring locations in simulation studies. In Section 5, we compare the proposed method to ProPrPCA in predicting exposure data at new locations under spatial misalignment, by reproducing the simulations and real data examples in \cite{vu2019probabilistic}. Finally, we discuss the implications of these results, our contributions, and potential extensions in Section 6. 

\section{Review of low-rank matrix completion (LRMC) \label{c3-review}}

\subsection{Problem formulation}

The optimization problem in (\ref{eq:tradl1}) can be referred to as a LR matrix approximation, or the ``fully observed" version of LRMC. When some entries of $\X$ are missing, the LR structure of $\X$ can be recovered by minimizing the residuals over only the observed indices, 
\begin{align}
\min_{\W} \hspace{0.15cm} \frac{1}{2} \left\{ \big\Vert P_{\Omega}(\X) - P_{\Omega}(\W) \big\Vert^2_F + \lambda \big\Vert \W \big\Vert_* \right\} , \label{eq:mcl1}
\end{align} 
where $\Omega$ denotes the set of observed indices in $\X$, and $[P_{\Omega}(\X)]_{ij} = X_{ij}$ if $(i, j) \in \Omega$ and zero otherwise. 

\subsection{Optimization}

\cite{mazumder2010spectral} prove that the ``fully observed" problem in (\ref{eq:tradl1}) has a closed-form solution $\hat{\W}$ that uses the \textit{soft-thresholding} operator,
$$\hat{\W} = \tilde{\U} S_{\lambda}{(\D)} \tilde{\V}^\tp.$$
Here $\tilde{\U} {\D} \tilde{\V}^\tp$ is the singular value decomposition (SVD) of $\X$, with $\D = \text{diag}\{\sigma_1, ..., \sigma_r\}$, with $\sigma_i$ being the $i-$th largest singular value of $\X$, and $r$ being the column rank of $\X$. We assume that the columns of $\X$ have been properly centered.  The soft-thresholding operator  \citep{donoho1995wavelet} is defined as $S_{\lambda}(\D) = \text{diag}\{(\sigma_1 - \lambda)_+, ..., (\sigma_r - \lambda)_+\}$, where $t_+ = \max(0, t)$. This result is closely related to PCA, in that, if PCA were to be applied onto $\X$, $\tilde{\V}$ would be returned as the loadings. 

\cite{mazumder2010spectral} propose an algorithm that uses proximal gradient descent \citep{rockafellar1976monotone} to solve (\ref{eq:mcl1}), which is given Table \ref{tab:lrmc}. A simple re-derivation of this LRMC algorithm is given in the Supplement. The algorithm consists of two major steps: a gradient descent update, and solving the proximal problem to (\ref{eq:mcl1}). In particular, the gradient update is simply filling in missing entries with the corresponding entries of the current estimate. The proximal problem turns out to be exactly the LR approximation problem (\ref{eq:tradl1}), which has a closed-form solution using soft-thresholding.

\section{The spatial matrix completion problem \label{c3-method}}

\subsection{Proposed optimization problem \label{c3-propose}}

The LRMC algorithm is a powerful tool to recover the LR structure of the data even though only a sampling of the entries is observed. Once the missing entries are filled, the LR structure, i.e. the PC scores, can be easily obtained by projecting the imputed data $\hat{\X}$ onto the direction of its first $q$ right singular vectors. However, under spatial misalignment, the ultimate goal is to produce accurate predictions at cohort locations where pollution data is unavailable. One potential approach is to employ a multi-stage procedure in these cohort studies: 1) imputation to fill in missing elements of the data matrix, 2) dimension reduction to obtain the LR structure of the data, and 3) spatial prediction to estimate these scores at locations of interest.  

In the second stage with dimension reduction, ideally we would like to identify principal directions such that the resulting PC scores would retain important characteristics and spatial structure. Having these spatial patterns, the PC scores could be predicted well at new locations in the spatial prediction stage. As a result, we propose the following convex optimization problem for the ``fully-observed" scenario,
\begin{align}
\min_{\M} \hspace{0.15cm} \left\{ \frac{1}{2} \big\Vert \X - \Z\M \big\Vert^2_F + \lambda \big\Vert \Z\M \big\Vert_*\right\}, \label{eq:smc-complete}
\end{align}
where $\Z \in \mathds{R}^{n \times k}$ contains geographic covariates used in the later prediction stage and thin-plate spline basis functions. The basis functions are included to capture any underlying spatial patterns that may not have been explained by other covariates.

When $\X$ has missing entries, we propose the following optimization problem
\begin{align}
\min_{\M} \hspace{0.15cm} \frac{1}{2} \left\{ \big\Vert P_{\Omega}(\X) - P_{\Omega}(\Z\M) \big\Vert^2_F + \lambda \big\Vert \Z\M \big\Vert_* \right\} , \label{eq:smc-missing}
\end{align} 

\subsection{Optimization}

First, we look at the complete data scenario with the proposed problem in (\ref{eq:smc-complete}). While $\M$ is the unknown quantity of the objective function, it is important to keep in mind that we are more interested in $\hat{\W} = \Z \hat{\M}$ where $\hat{\M}$ is the optimal solution for (\ref{eq:smc-complete}). This quantity is the LR approximation of $\X$.  We give the closed-form solution of $\hat{\W}$ in the following lemma, with the detailed proof in the Supplement.

\noindent \textbf{Lemma.} \textit{If $(\Z^\tp \Z)^{-1}$ exists, then the approximation $\hat{\W} = \Z \hat{\M}$ of $\X$, where $\hat{\M}$ is the optimal solution for (\ref{eq:smc-complete}), has a closed-form expression, $\hat{\W} = \tilde{\U} S_{\lambda}({\D} )\tilde{\V}^\tp$, where $\tilde{\U} {\D} \tilde{\V}^\tp$ is SVD of $\tilde{\X} = \Z(\Z^\tp \Z)^{-1}\Z^\tp \X$.}

When $\X$ has missing entries, we prove that $\hat{\W} = \Z \hat{\M}$, where $\hat{\M}$ is the optimizer of (\ref{eq:smc-missing}), can be derived via a proximal algorithm similar to the LRMC algorithm. The steps are summarized in Table \ref{tab:smc}. It turns out that our algorithm is similar to the LRMC algorithm, with the exception of the insertion of an additional projection step involving $\Z$.  Derivation and proof are given in the Supplement. We refer to this as the \textbf{S}patial \textbf{M}atrix \textbf{C}ompletion (SMC) algorithm.

\subsection{Connection with existing methods \label{c3-connection}}

Another direct approach to induce spatial patterns into the PC scores was the PredPCA algorithm proposed by \cite{jandarov2017novel}. The PredPCA algorithm also employs the same matrix $\Z$ of covariates and spline basis functions in its objective function. This algorithm uses a biconvex formulation of PCA where the PCs are estimated sequentially, 
$$\min_{\alp, \bv}   \bigg\Vert \X - \left( \frac{\Z \alp}{\Vert \Z \alp \Vert_2} \right) \bv^\tp \bigg\Vert^2_F ,$$
where $\alp \in \mathds{R}^k$ and $\bv \in \mathds{R}^p$. Here the algorithm estimates one PC at a time. The quantity ${\Z \alp}/{\Vert \Z \alp \Vert_2}$ plays the role of the PC score, while $\bv$ is the loading. PredPCA directly imposes a spatial structure on the score via $\Z$. By doing so, PredPCA essentially aims to recover the best rank-1 approximation of $\X$ that has spatial structure embedded within the left singular vectors. Thus the objective function of PredPCA can be rewritten as an L0 problem with spatial constraints.  Heuristically, this is similar to what SMC aims to achieve. The fundamental difference is that SMC utilizes a nuclear-norm penalty, while PredPCA resembles a spatial optimization problem with an L0 penalty. When some data are missing, such reformulated version of PredPCA can potentially be solved using hard-thresholding \citep{mazumder2010spectral}. However, unlike SMC, the lack of convexity means that convergence to an optimal solution and corresponding theoretical properties are not guaranteed. 

In its current form, the sequential estimation and the lack of a mechanism to handle missing data are rather unsatisfying. The probabilistic version, ProPrPCA \citep{vu2019probabilistic}, provides a better alternative to PredPCA when missing data are present, as imputation and dimension reduction are handled simultaneously. However, ProPrPCA also involves a series of rank-1 approximations, and requires longer computational time overall due to numerical optimizations of spatial parameters and likelihood-based imputations for missing observations. 

In later sections, we compare SMC to both LRMC, in terms of recovering missing entries at monitoring locations, and ProPrPCA, in terms of predicting exposure data at new locations of interest. 

\section{Low-rank matrix recovery in simulation studies: A comparison between SMC and LRMC \label{c3-sim-recovery}}

\subsection{Setups}

We first conduct two sets of simulation studies to compare the performance of LRMC and SMC in \textit{recovering missing entries}. We generate spatially-correlated multivariate data on a dense $100\times100$ grid ($N=10,000$), using the following formula
$$\X = \left( \R_o \B_o + \R_u \B_u + \bS  \right)\V^\tp + \E. $$
Here $\R_o$ and $\R_u$ represent the observed and unmeasured covariates, respectively, in addition to the data of interest $\X \in \mathds{R}^{N \times p}$. In practice, these covariates often include, but not limited to, Geographic Information System (GIS) variables. $\bS$ is a mean-zero stationary structure with exponential spatial covariance, while $\E$ include independent white noises. The loadings $\V$ represent the mapping from the LR $q-$dimensional space to the higher $p-$dimensional space of the data. The specific simulation parameters can be found in the Supplement. The first set is a toy simulation study with $q=1$ and $p=4$, while the second set of simulations has $q = 3$ and $p=12$. We consider four scenarios defined by the presence of unmeasured covariates ($\B_u = \bze$ or $\B_u \neq \bze$) and spatial correlations ($\bS = \bze$ or $\bS \neq \bze$). 

In each simulation, we randomly choose 400 locations, and omit data entries at random. The level of missing data completely at random (MCAR) for each feature in $\X$ varies from 5\% to 40\%, with a 5\% increment. We use LRMC and SMC to recover the missing entries and evalue the mean squared errors (MSE) of these entries. The design matrix $\Z$ used in SMC includes the observed covariates $\R_o$ and thin-plate spline basis functions to capture other spatial variability. 

While we focus on matrix recovery in this section, in each simulation, we also pick 100 new locations at random. Under spatial misalignment, the scientific interest lies in the data at these new locations, however, they are 100\% missing. Thus accurate prediction of data at these locations using the ``observed" locations is necessary. As SMC assumes a mean model for the  LR structure of the data, this method is capable of producing an estimate for the data at new locations where no observation on any feature of $\X$ is available, while LRMC cannot. Thus, we also evaluate the MSE of SMC at these new locations.  

\subsection{Results}

\begin{figure}[H]
	\begin{center}
		\includegraphics[width=6in]{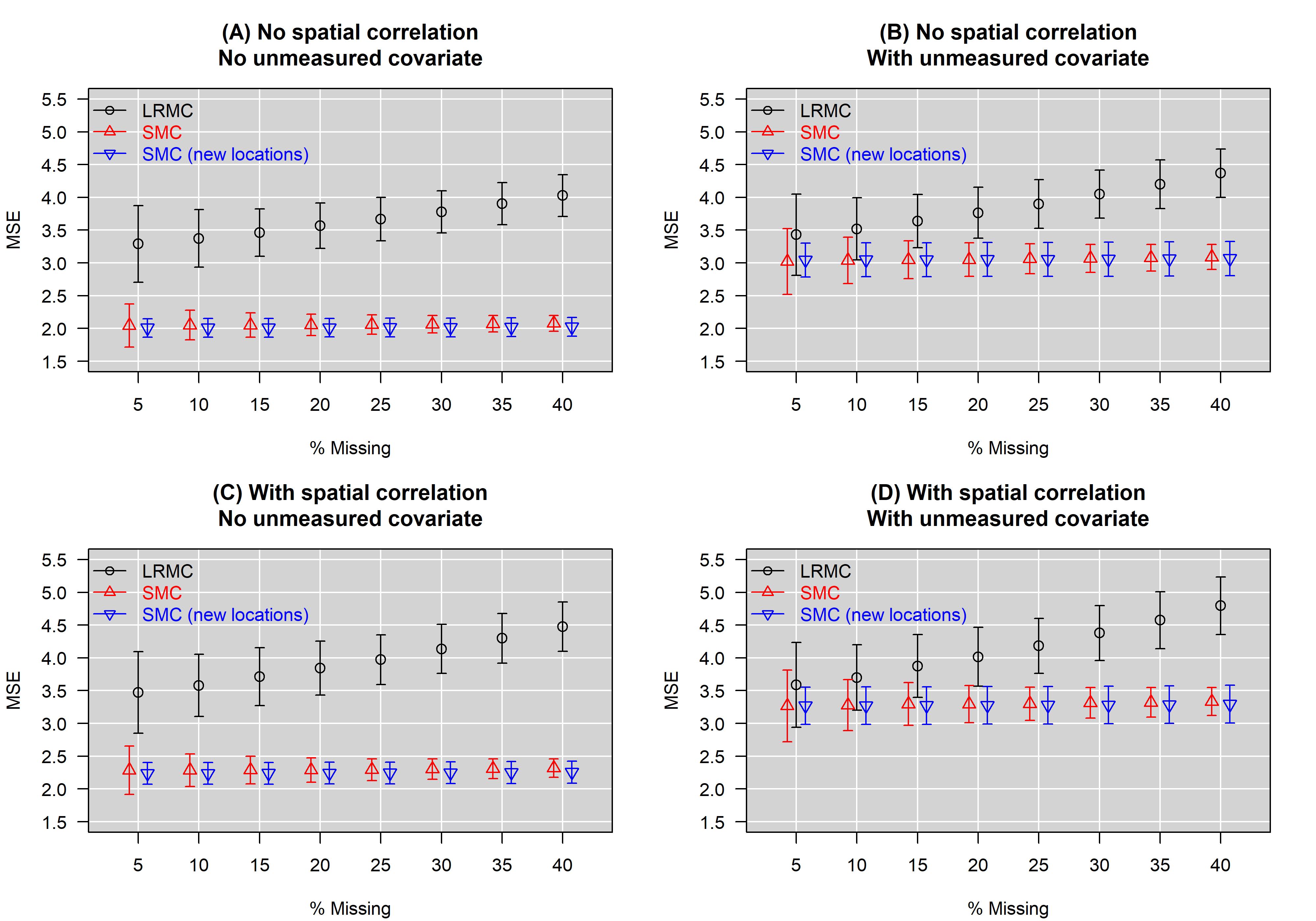}
	\end{center}
	\caption{\doublespacing Mean (SD) for MSE at missing entries (LRMC in black, and SMC in red), and MSE for all entries at new locations (SMC in blue) across 1000 simulations for various level of MCAR and different data scenarios in toy simulations, where four features ($p = 4$) are generated based on a rank-1 structure ($q = 1$).}
		\label{fig:simlow}
\end{figure}

Figure \ref{fig:simlow} shows the results of the toy simulations in which the data with four features ($p = 4$) are generated based on a rank-1 structure ($q = 1$). Overall, the mean MSE values for LRMC are always higher than those of SMC in these simulations. As the amount of missing data increases, mean MSE goes up for LRMC, but remains relatively the same for SMC even at MCAR level as high as 40\%. It is notable that the variability of SMC results (in red) is the highest at 5\% MCAR and decreases with more missing data. This is possibly due to the algorithm's greedy attempt to use rich patterns learned within observed entries to recover a much smaller number of missing entries. 

The advantage of SMC over LRMC is most apparent when there is no unmeasured covariate, i.e. $\B_u = \bze$. When there are unmeasured covariates (panels B and D), the unobserved component, $\R_u \B_u \V^\tp$, is simply part of the LR structure data, which LRMC algorithm can leverage, and thus its performance is not substantially impacted. With a large amount of unmeasured LR structure, it is possible that LRMC may outperform SMC at low level of missing data. However, even in these scenarios, the performance of SMC stays consistent as the amount of missing data increases. 

In scenarios where there is truly no spatial correlation (panels A and B), SMC still outperforms LRMC. It is because SMC utilizes relevant covariates $\R_o$ that are predictable of the simulated data. Finally, SMC is capable of giving reasonable estimates of the data at new locations with no observed feature (results in blue), which LRMC cannot. This is an important implication in the context of spatial misalignment that is commonly seen in air pollution cohort studies. 

The results from the higher-dimensional simulations in which the data generated with 12 features are included in the Supplement. We observe trends similar to these results from the toy simulations. 

\section{Predictive performance under spatial misalignment: A comparison between SMC and ProPrPCA \label{c3-reproduce} }

As alluded to in Section \ref{c3-propose}, in health-pollution cohort studies, the goal is not to recover missing data at monitoring locations. However, accurate predictions of the total mass and chemical profile at cohort locations, where pollution data are not readily available, are often required prior to health-pollution analysis. A straightforward solution is a multi-stage approach with imputation (to fill in missing monitoring data if any), dimension reduction (to identify important pollutant profile and also reduce computational burden of later step), and spatial prediction (estimate exposures at unobserved locations). As discussed in Section \ref{c3-connection}, existing methods PCA and PredPCA require complete data as input. LRMC can be used in the imputation step, however, it does not leverage important information from relevant covariates that are often available in environmental studies, such as GIS variables. 

ProPrPCA \citep{vu2019probabilistic} provides likelihood-based imputation simultaneously within the dimension reduction step. This method also aims to produce a lower-dimensional representation of the original data that can be easily predicted at new locations. Our proposed SMC model offers an alternative approach, with similar goals to ProPrPCA, but using convex optimization. Thus, in this Section we reproduce the simulations and data application in \cite{vu2019probabilistic}, to compare SMC directly to ProPrPCA, PredPCA, and traditional PCA. 

\subsection{Simulation studies}

The full details of the original simulations are described in \cite{vu2019probabilistic}. In summary, these simulations include $p=15$ pollutant surfaces spanning a dense $100\times100$ grid and generated based on three underlying PC scores. These scores have various levels of variance contribution to the data overall and \textit{spatial predictability}, i.e. how well these scores can be predicted at new locations using external covariates. In the first scenario where the orders of spatial predictability and variance contribution to the data are the same, all competing methods are expected to perform equally well when there is no missing training data. The second scenario is where the most spatially predictable score did not contribute the most variability to the data. We also consider various settings in which the training data are either complete, missing completely at random (MCAR), or missing at random (MAR), where the missing patterns are associated with external covariates. 

In each of the 1000 simulations, 400 training locations and 100 test locations are chosen at random. We implement the multi-stage procedure and evaluate these competing dimension reduction methods based on their performance in the spatial prediction stage with universal kriging. When training data are complete, we compare the ``fully-observed" version of SMC to PCA, PredPCA, and ProPrPCA. When the training data are either MCAR or MAR, LRMC is used in the imputation step prior to PCA and PredPCA, while neither ProPrPCA nor SMC requires this extra step. Specially, we compare the predicted scores, estimated at test locations by using universal kriging and the training scores obtained from each dimension reduction method, with the ``true" scores, derived by projection of test data. 

Table \ref{tab:s1} provides both the prediction R$^2$ used in \cite{vu2019probabilistic}, which reflects the correlation between the predicted score and true score, as well as the median MSE for each PC and for the overall reconstructed data matrix at test locations in the first simulation scenario. As expected, all methods perform equally well when the training data are complete. Under MAR, data among the first 5 pollutants are more likely to be missing at locations with extreme geographic covariates values. As discussed in the original paper, due to the data generating mechanism, this setup eventually leads to the decreases in median R$^2$'s for PC1 but slight increases for PC2. While the overall performance decreases under MAR, SMC gives the best result for PC1 (median R$^2$ = 0.74), followed by ProPrPCA (median R$^2$ = 0.69). This is in exchange for a slightly better performance for PC2 by ProPrPCA. 

\begin{figure}[H] 
	\centering
	\includegraphics[width=6.5in]{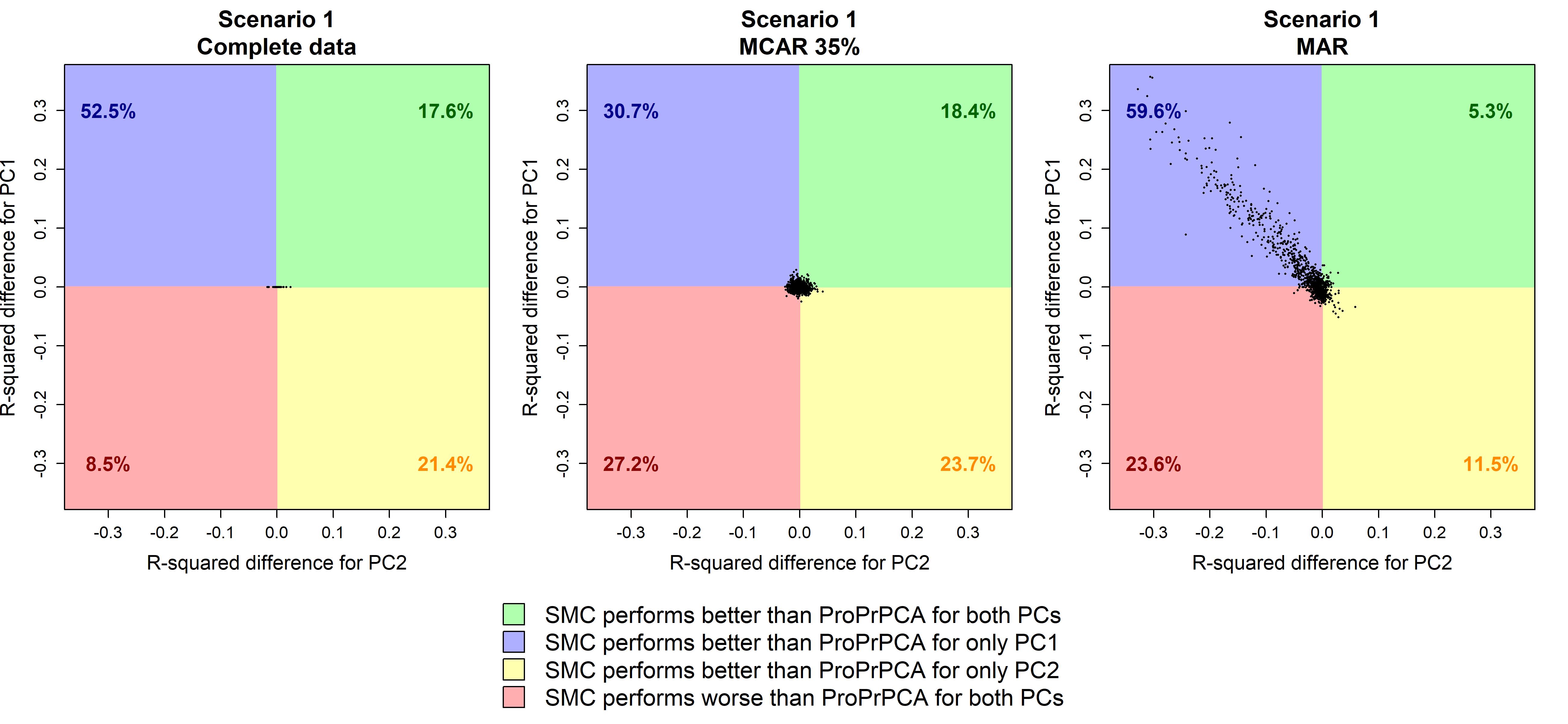}
	\caption{\doublespacing Difference in R$^2$ between SMC and ProPrPCA for scenario 1. Each dot represents result from one simulation. Percentages indicate the proportion out of 1,000 simulations.} 
	\label{fig:s1-pair-smc-spline}
\end{figure}

The differences between SMC and ProPrPCA are further examined in Figure \ref{fig:s1-pair-smc-spline}, where we examine scatter plot of prediction R$^2$ across all simulations. The discrepancy is negligible with complete data or MCAR 35\%. Under MAR, ProPrPCA shows some advantage by beating SMC for both PCs in 23.6\% of the simulations (bottom-left quadrant), compared to just 5.3\% of the time for SMC (top-right quadrant). However, the magnitudes of differences in these regions are relatively small. In 59.6\% of the simulations, SMC is better at predicting PC1 with differences ranging up to 0.3 in R$^2$. Similar comparison between SMC and PredPCA are shown in the Supplement.

Corresponding results for the second simulation scenario are shown in the Supplement. Overall, we find that SMC produces results that are on par with ProPrPCA in simulations with various levels of missing data. 

\subsection{Assessment of computational burden}

\begin{figure}[H]
	\begin{center}
		\includegraphics[width=5in]{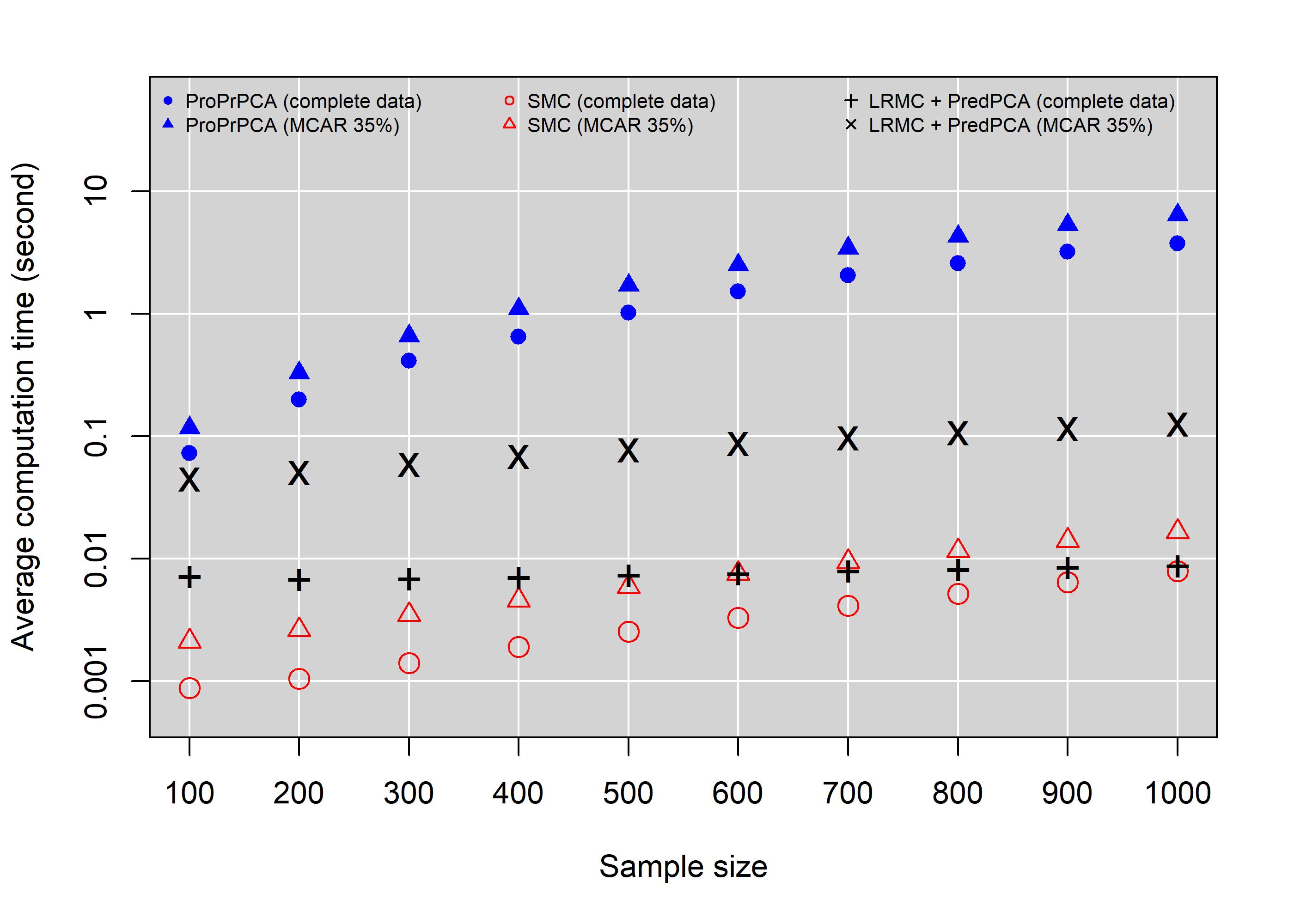}
	\end{center}
	\caption{\doublespacing Computation time (average over 1,000 simulations) of ProPrPCA, PredPCA, and SMC with complete and MCAR 35\% missing data by sample size. Under missing data scenarios, LRMC is applied prior to implementing PredPCA. 
		\label{fig:time}}
\end{figure}

In addition, we find SMC to be more computationally efficient as the sample size increases, as shown in Figure \ref{fig:time}. Even though ProPrPCA only takes about 10 seconds for $n=1000$ on average, due to complex likelihood-based algebras, ProPrPCA is not as efficient as methods that leverage convex optimization. Even with 2 separate steps of imputation and dimension reduction, LRMC+PredPCA is faster at roughly 0.1 second run time on average. SMC proves to be the most efficient at approximately 0.01 second on average with the same sample size. The increase in computational burden jumps from 0.1 ($n=100$) to 10 seconds ($n=1000$) on average for ProPrPCA. In contrast, the burden only increases 10 fold from 0.001 to 0.01 for SMC. Although 10 seconds may not seem to be a big deal in practice for one dataset, this has a potential implication for larger datasets in practice that span multiple years or with complicated missing patterns that can complicate the likelihood calculations used in ProPrPCA. Meanwhile SMC has the potential to scale and be further optimized thanks to using similar structure to LRMC.  

\subsection{Data application}

Next we apply SMC to the same data application used in \cite{vu2019probabilistic}, which were collected nationally by the Chemical Speciation Network (CSN), a sub-network of the Air Quality System network of monitors managed by the Environmental Protection Agency. Data are available for 21 component of PM$_{2.5}$. GIS covariates are provided by the MESA Air Team at the University of Washington. For consistency with all previous literature \citep{vu2019probabilistic, jandarov2017novel, keller2017covariate}, we use the 2010 CSN data, which after data processing described in \cite{vu2019probabilistic}, consists of 221 CSN sites, only 130 sites of which have complete data for all 21 components (overall 32.1\% missing data). We additionally look into similar data collected in 2011, which includes 208 sites in total, only 128 sites of which have complete data (overall 30.1\% missing data).

We first evaluate how the pollutant profile varies across dimension reduction methods. We also assess how the estimated loadings and corresponding PC scores change with different methods and whether all sites or just complete sites are used. The results are given in the Supplement. Overall, we find that the estimated loadings are noticeably different when including sites with missing data. SMC gives results that shares similarities with both PredPCA and ProPrPCA. 

Next we evaluate the predictive performance via leave-one-site-out cross-validation. In each iteration, we leave one site with complete data on all 21 components as test data, while performing dimension reduction and building spatial prediction model based on training data. For a fair comparison, we use the same universal kring model with exponential covariance structure for the spatial prediction step. The training data in each iteration may consist of either the remaining complete site (``complete" training data), or all remaining site (``full" training data). Similar to \cite{vu2019probabilistic}, we focus on the prediction for each PC instead of reconstructing the data matrix, because in practice, these PCs would represent the pollutant profile, which could be used as an effect modifier in a model investigating the associations between total mass of particulate matter and health outcomes. 

The results for 2010 CSN data are shown in Table \ref{tab:cross-validation-2010}. When using only complete sites for training data, PCA has acceptable performance for PC2 and PC3, however, gives poor results for PC1 (R$^2 = 0.24$). Meanwhile, the other three methods give the same results and better R$^2$ for both PC1 and PC3. When training data consist of all remaining sites, SMC is better than PredPCA overall for all PCs. SMC has the same performance as ProPrPCA for PC1, but there is a slight trade-off in performance between PC2 and PC3. SMC has a substantially higher R$^2$ for PC2 compared to ProPrPCA, but worse performance for PC3. These results may be due to the switching between PC2 and PC3 observed in the pollutant profile, as shown in the Supplement. The overall performance for 2011 data, as given in the Supplement, is better than 2010 across all methods. When using all available sites, the performance of SMC is more similar to ProPrPCA in this dataset compared to 2010 data. 

For a fair comparison across all methods, we use the same pre-specified spatial information encoded in $\Z$ and the same universal kriging model in the spatial prediction stage for all PCs. While simplifying the comparisons, using the same $\Z$ matrix to model all PCs may not be effective, and the covariance structure might have not been correctly specified in these data applications. Potential solutions, which are beyond the scope of this paper, include an adaptive selection of features to be included in $\Z$ \citep{bose2018adaptive}, or more sophisticated non-stationary models for spatial prediction instead of universal kriging. 

\section{Discussion \label{c3-discussion}}

In this paper, we focus on problems arising in health-pollution cohort studies, in which multi-pollutant data are often spatially misaligned and have a large number of missing observations. The ultimate goal is to develop a dimension reduction technique that is similar to PCA but able to  leverage spatial structure and external information to obtain accurate predictions at new locations of interest. The scientific motivation includes the ability to characterize the pollutant profile across locations, and to identify effect modifiers for the health-pollution associations of interest. For example, many studies on fine particulate matter (PM$_{2.5}$) have shown evidence that the associations between health outcomes and PM$_{2.5}$ total mass can be significantly modified by the PM$_{2.5}$ chemical composition \citep{krall2013short, zanobetti2014health, kioumourtzoglou2015pm2, wang2017long, keller2018pollutant}. 

We formulate a convex optimization problem based on the existing idea of LRMC with nuclear-norm penalization. We show that a closed-form solution exists when the original data are fully observed. In addition, we also derive a proximal algorithm to solve the problem when some elements of the  data are missing. In simulations with generated spatial data, we show that SMC outperforms LRMC in recovering missing data entries. We also illustrate how SMC can produce reliable estimates of the entire pollutant profile at new locations while LRMC simply cannot. In addition, we show the merits of SMC under spatial misalignment typically seen in air pollution cohort studies by reproducing and comparing to simulated results and data applications given in \cite{vu2019probabilistic}. 

A slight complication of SMC compared to other methods is that it requires the penalty parameter $\lambda$. In our current algorithm, the choice of $\lambda$ is based on a small grid search to reach the desired rank $q$ of the low-rank approximation. However, the grid search does not have a major impact on computation time, as shown in simulation results. Computation time can also be shortened using warm starts \citep{mazumder2010spectral}.

Under the setting of spatially misaligned and multivariate air pollution data, SMC can produce LR structure with spatial patterns and impute for missing data with considerations of external geographic and spatial information, as illustrated in Table \ref{tab:compare}. SMC is also able to estimate all PCs simultaneously, whereas the ProPrPCA model requires the PCs to be estimated sequentially. Under SMC, estimated PCs are guaranteed to be orthogonal, and thus considered to be uncorrelated, which is one of the desirable properties of PCA. It is important to note that for ProPrPCA, when data is missing, the parameter estimation and data imputation are separate. The loadings are estimated based only on the observed data. The data is then imputed with consideration of $\Z$, and projected onto the directions of the loadings to derive the PC scores. One can practically use different $\Z$ matrices for the estimation and imputation procedures. This can potentially be more beneficial and more accurate, particularly when there is reasonable evidence to believe that the missing mechanism only depends on a subset of covariates and spline terms included in $\Z$. Meanwhile, imputation and dimension reduction are essentially intertwined in the SMC algorithm, and there is no flexibility in modifying information used for imputing data only. The results discussed in Section \ref{c3-reproduce} show that ProPrPCA consistently produces better results, although SMC follows very closely. The SMC algorithm offers a faster, more compact and elegant alternative to ProPrPCA. 

In its current form proposed by \cite{jandarov2017novel}, an imputation step is required prior to PredPCA. LRMC is a useful method to fill in the missing data, but it does not account for spatial structure while imputing the data. Using LRMC prior to PredPCA may distort the underlying structure of the data even before dimension reduction, and thus worsen the predictive performance. 

In recent literature, LRMC has been employed in various problems involving spatially correlated data. For example, LRMC is used in video denoising \citep{ji2010robust}, seismic data reconstruction \citep{yang2013seismic}, and imaging recovery \citep{shin2014calibrationless}. \cite{cabral2014matrix} tackles the problem of multi-label image classification by extending LRMC with different loss functions to reflect the correct constraints of imaging data. \cite{xie2017recover} develops a two-phase matrix-completion-based procedure with spatial and temporal considerations to recover corrupted weather data. These methods are intriguing and work well for the purpose of handling correlated missing data. However, none of these approaches directly impose spatial patterns into the the LR structure of the data. To the best of our knowledge, no other method has utilize the LRMC framework in such a way that is relevant to spatially-misaligned multi-pollutant data. 

While initially focusing on health-pollution cohort studies, our proposed framework can be applicable to other fields where spatial misalignment motivates a separate exposure model. In such cases, the design matrix $\Z$ can be modified to incorporate whatever covariates are necessary, with spline terms that represents various structures not limited to just spatial correlations. Future work includes potential extension to misaligned spatio-temporal data. 

\section*{Supporting information}

Additional information and supporting material for this article is available online at the journal's website. Data used in this paper are available upon request through the MESA Air team at the University of Washington.

\section*{Acknowledgment}

We thank the University of Washington MESA Air team for providing the data to reproduce the results shown in \cite{vu2019probabilistic}. The data are available upon request by contacting the University Washington MESA Air team directly. 

Dr. Szpiro received funding as part of the NIH/NIEHS grants 1R21ES024894 and 5R01ES026246. The content of this paper is solely the responsibility of the authors, and does not necessarily represent the official views of the National Institutes of Health/National Institute of Environmental Health Sciences. 

\bibliographystyle{apalike}
\bibliography{VuPT-SMC}

\clearpage


\begin{figure}[H]
	\begin{center}
		\includegraphics[width=3.5in]{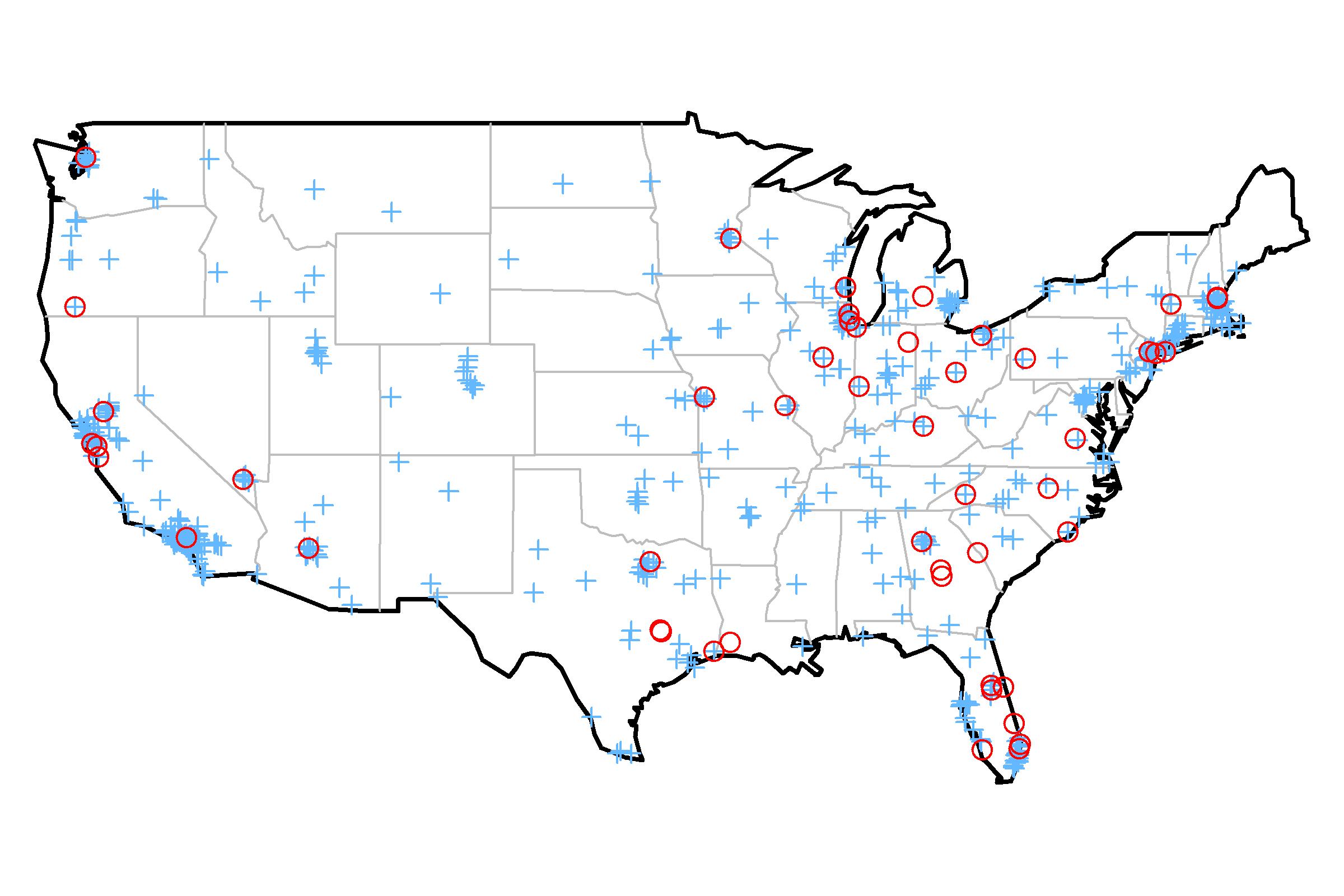}
	\end{center}
	\caption{\doublespacing Illustration of spatial misalignment in health-pollutant cohort studies. The red dots represent the locations of fixed monitoring sites where pollutant data is available. The blue crosses mark the locations of study participants at which health outcome is available.} 
		\label{fig:spatialmisalignment}
\end{figure}

\begin{figure}[H]
	\begin{center}
		\includegraphics[width=6in]{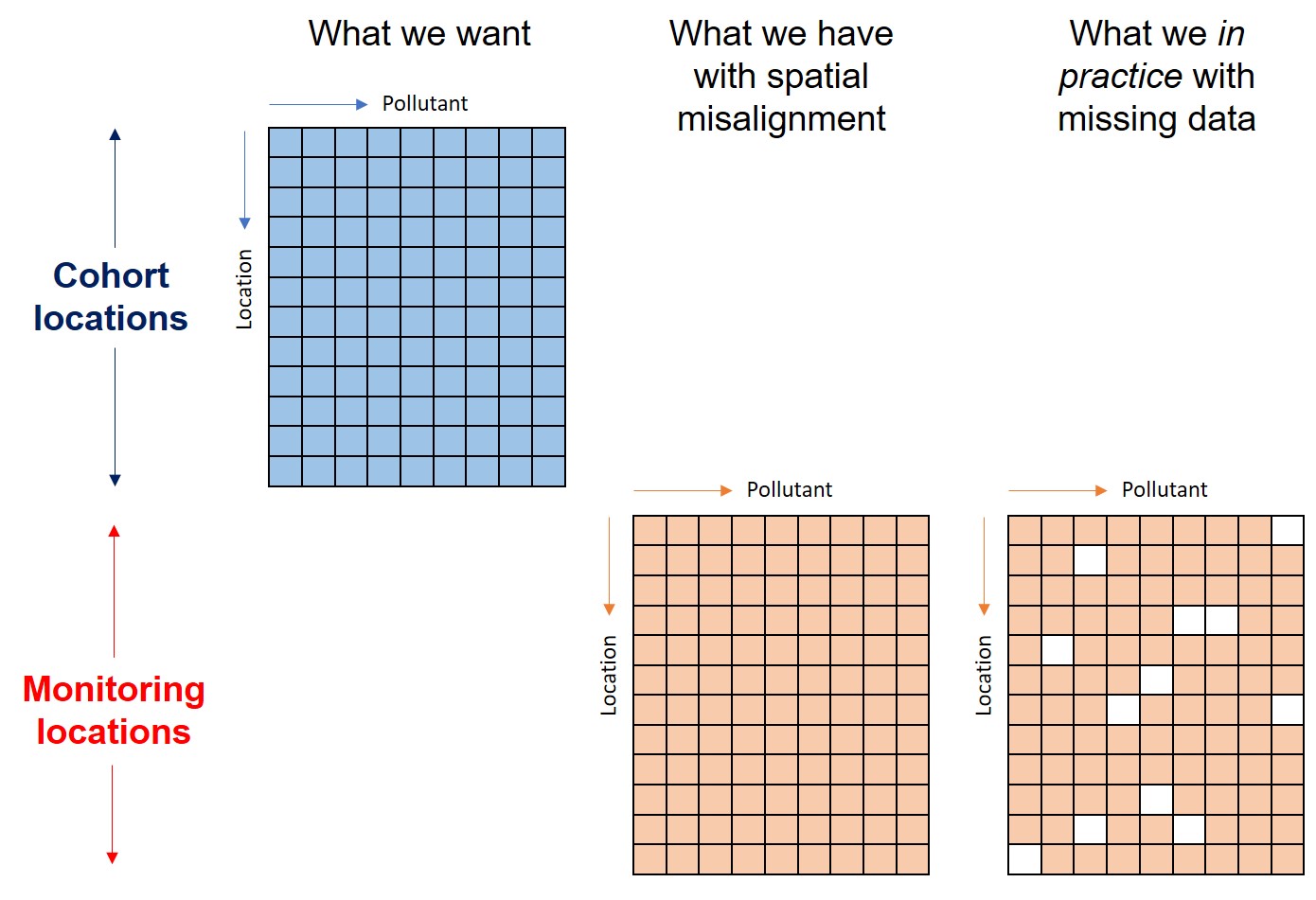}
	\end{center}
	\caption{\doublespacing Illustration of the usual challenges in practice with using spatially misaligned and multivariate exposure data in health-pollution cohort studies.} 
	\label{fig:challenge}
\end{figure}

\begin{table}[H]
	\begin{center}
		\caption{ Algorithm 1 - LRMC adapted from \cite{mazumder2010spectral}} 
		\label{tab:lrmc}
		\begin{tabular}{@{}*{1}{p{\textwidth}@{}}}
			\toprule
			\textbf{Input} $\X$, $q$, $\lambda$, and $t_{max}$ \\
			\hspace{0.5cm} \textbf{Initialize} $\W^{(0)} = \bze$, $t = 1$ \\
			\hspace{1cm} \textbf{while} not converged or $t < t_{max}$ \textbf{do} \\
			\hspace{1.5cm} $\tilde{\W}^{(t)} \leftarrow P_{\Omega}(\X) + P^{\perp}_{\Omega}(\W^{(t)})$, where $P^{\perp}_{\Omega}(\W^{(t)}) = \W^{(t)} - P_{\Omega}(\W^{(t)})$ \\
			\hspace{1.5cm} ${\W}^{(t+1)} \leftarrow \tilde{\U} S_{\lambda}({\tilde{\D}}) \tilde{\V}^\tp$, where $\tilde{\U} \tilde{\D} \tilde{\V}^\tp$ is the SVD of $\tilde{\W}^{(t)}$ \\
			\hspace{1.5cm} $t \leftarrow t+1$ \\
			\hspace{1cm} \textbf{end while} \\
			\textbf{Output} $\hat{\W} = \W^{(t)}$, $\hat{\X} = P_{\Omega}(\X) + P^{\perp}_{\Omega}(\hat{\W})$ \\
			\bottomrule
		\end{tabular}
	\end{center}
\end{table}

\begin{table}[H]
	\begin{center}
		\caption{ Algorithm 2 - Spatial matrix completion (SMC) } 
		\label{tab:smc}
		\begin{tabular}{@{}*{1}{p{\textwidth}@{}}}
			\toprule
			\textbf{Input} $\X$, $\Z$, $q$, $\lambda$, and $t_{max}$ \\
			\hspace{0.5cm} \textbf{Initialize} $\W^{(0)} = \bze$, $t = 1$ \\
			\hspace{1cm} \textbf{while} not converged or $t < t_{max}$ \textbf{do} \\
			\hspace{1.5cm} $\breve{\W}^{(t)} \leftarrow P_{\Omega}(\X) + P^{\perp}_{\Omega}(\W^{(t)})$ \\ 
			\hspace{1.5cm} $\tilde{\W}^{(t)} \leftarrow \Z (\Z^\tp\Z)^{-1} \Z^\tp \breve{\W}^{(t)} $ \\ 
			\hspace{1.5cm} ${\W}^{(t+1)} \leftarrow \tilde{\U} S_{\lambda}({\tilde{\D}}) \tilde{\V}^\tp$, where $\tilde{\U} \tilde{\D} \tilde{\V}^\tp$ is the SVD of $\tilde{\W}^{(t)}$ \\
			\hspace{1.5cm} $t \leftarrow t+1$ \\
			\hspace{1cm} \textbf{end while} \\
			\textbf{Output} $\hat{\W} = \W^{(t)}$, $\hat{\X} = P_{\Omega}(\X) + P^{\perp}_{\Omega}(\hat{\W})$ \\
			\bottomrule
		\end{tabular}
	\end{center}
\end{table}

\begin{table}[H] 
	\caption{\doublespacing The predictive performance for simulation scenario 1 is evaluated on the test data for the four competing methods. The results are displayed as median R$^2$ and median MSE (in brackets) for the first two PCs, and median MSE for the entire data matrix reconstructed at test locations, across 1000 simulations. Variability across simulations is similar for all methods and hence not shown in this table. Under missing data scenarios, LRMC is used prior to either PCA or PredPCA. } 
	\label{tab:s1}
	\centering
	\vspace{0.25cm}
	\begin{tabular}{lrrr}
		\hline
		 & \textbf{Complete} & \textbf{MCAR 35\%} & \textbf{MAR} \\ \hline
		
		\textbf{PC1} - median R$^2$ [median MSE] & & & \\
		\hspace{0.5cm} PCA &  0.83 [1.84] & 0.80 [2.10] & 0.61 [3.35] \\
		\hspace{0.5cm} PredPCA & 0.84 [1.76] & 0.81 [2.03] & 0.63 [3.22] \\
		\hspace{0.5cm} ProPrPCA & 0.84 [1.76] & 0.83 [1.79] & 0.69 [2.88] \\
		\hspace{0.5cm} SMC &  0.84 [1.76] & 0.83 [1.80] & 0.74 [2.70] \\ \hline
		\textbf{PC2} - median R$^2$ [median MSE] & & & \\
		\hspace{0.5cm} PCA & 0.60 [3.31] & 0.58 [3.45] & 0.67 [3.31] \\
		\hspace{0.5cm} PredPCA & 0.60 [3.32] & 0.58 [3.45] & 0.68 [3.22] \\
		\hspace{0.5cm} ProPrPCA & 0.60 [3.31] & 0.60 [3.32] & 0.68 [2.97] \\
		\hspace{0.5cm} SMC & 0.60 [3.31] & 0.60 [3.33] & 0.63 [3.18] \\ \hline
		\textbf{Reconstructed test data} - median MSE & & & \\
		\hspace{0.5cm} PCA & 1.54 & 1.57 & 1.66 \\      
		\hspace{0.5cm} PredPCA & 1.53 & 1.56 & 1.63 \\      
		\hspace{0.5cm} ProPrPCA & 1.53 & 1.54 & 1.58 \\      
		\hspace{0.5cm} SMC & 1.53 & 1.54 & 1.59 \\ \hline    
	\end{tabular}

\end{table}

\begin{table}[H]
	\centering
	\caption{\doublespacing Prediction results from leave-one-site-out cross-validation on 2010 CSN data, using training data with either complete sites only or all available data, are reported as prediction R$^2$ and MSE (in brackets) for each PC and the overall reconstructed data matrix. LRMC is used prior to PCA or PredPCA when training set has missing data. Only sites with complete PM$_{2.5}$ component data are used as test data. }
	\begin{tabular}{lccc} 
		\hline
		 & \textbf{PC1} & \textbf{PC2} & \textbf{PC3}   \\ \hline
		\textbf{Training data with complete sites only} - R$^2$ [MSE] & & &  \\ 
		\hspace{0.5cm} PCA & 0.24 [0.91] & 0.51 [0.33] & 0.51 [0.26]  \\ 
		\hspace{0.5cm} PredPCA & 0.52 [0.47] & 0.44 [0.34] & 0.62 [0.22]  \\ 
		\hspace{0.5cm} ProPrPCA & 0.52 [0.47] & 0.44 [0.33] & 0.62 [0.22]  \\ 
		\hspace{0.5cm} SMC & 0.52 [0.47] & 0.44 [0.33] & 0.62 [0.22] \\  \hline
		\textbf{Training data with all available sites}  - R$^2$ [MSE] & & &  \\ 
		\hspace{0.5cm} PCA & 0.32 [0.52] & 0.44 [0.39] & 0.52 [0.25]  \\ 
		\hspace{0.5cm} PredPCA & 0.54 [0.43] & 0.53 [0.20] & 0.45 [0.39]   \\ 
		\hspace{0.5cm} ProPrPCA & 0.57 [0.41] & 0.35 [0.32] & 0.69 [0.21]  \\ 
		\hspace{0.5cm} SMC & 0.57 [0.40] & 0.51 [0.21] & 0.48 [0.38] \\
		\hline
	\end{tabular}
	\label{tab:cross-validation-2010}
\end{table}

\begin{table}[H] 
	\caption{\doublespacing Evaluation of different approaches with imputation and dimension reduction in the setting of spatially misaligned and multivariate air pollution data with missing observations.} 
	\label{tab:compare}
	\centering
	\begin{tabular}{lcccc}
		\toprule
		& LRMC +  & LRMC + & & \\
		& PCA & PredPCA & ProPrPCA & SMC \\
		\midrule
		Induce spatial patterns in PC scores & & $\checkmark$ & $\checkmark$ & $\checkmark$  \\
		Impute data with spatial consideration & & & $\checkmark$ & $\checkmark$ \\
		Estimate all PCs simultaneously & $\checkmark$ & & & $\checkmark$ \\
		\bottomrule
	\end{tabular}
\end{table}

\end{document}